\documentclass[a4paper]{spie}  %>>> use this instead for A4 paper
\usepackage[dvipdfmx]{graphicx}

\title{Imaging and spectral performance of\\ CdTe double-sided strip detectors for the Hard X-ray Imager onboard ASTRO-H} 

\author{Kouichi Hagino\supit{a}\supit{b}, Hirokazu Odaka\supit{a}, Goro Sato\supit{a}, Shin Watanabe\supit{a}\supit{b}, Shin'ichiro Takeda\supit{a}, Motohide Kokubun\supit{a}, 
Taro Fukuyama\supit{a}\supit{b}, Shinya Saito\supit{a}\supit{b}, 
Tamotsu Sato\supit{a}\supit{b}, Yuto Ichinohe\supit{a}\supit{b}, Tadayuki Takahashi\supit{a}\supit{b},
Toshio Nakano\supit{b}, Kazuhiro Nakazawa\supit{b}, Kazuo Makishima\supit{b}, \\
Hiroyasu Tajima\supit{c},
Takaaki Tanaka\supit{d},
Kazunori Ishibashi\supit{e}, Takuya Miyazawa\supit{e}, Michito Sakai\supit{e}, Karin Sakanobe\supit{e}, Hiroyoshi Kato\supit{e}, Shunya Takizawa\supit{e},
Kentaro Uesugi\supit{f}
\skiplinehalf
\supit{a}ISAS/JAXA, 3-1-1 Yoshinodai Chuo-ku Sagamihara Kanagawa 252-5210, Japan; \\
\supit{b}Department of Physics, The University of Tokyo, 7-3-1 Hongo Bunkyo-ku Tokyo 113-0033, Japan; \\
\supit{c}STEL, Nagoya University, Furo-cho Chikusa-ku Nagoya 464-8601, Japan; \\
\supit{d}Department of Physics, Kyoto University, Kitashirakawa-Oiwakecho Sakyo-ku Kyoto 606-8502, Japan; \\
\supit{e}Department of Physics, Nagoya University, Furo-cho Chikusa-ku Nagoya 464-8602, Japan; \\
\supit{f}Japan Synchrotron Radiation Research Institute, 1-1-1 Kouto Sayo-cho Sayo-gun Hyogo 679-5198, Japan
}

\authorinfo{Further author information: (Send correspondence to K.Hagino) K.Hagino: E-mail: hagino@astro.isas.jaxa.jp}
\begin{document} 
\maketitle 

%%%%%%%%%%%%%%%%%%%%%%%%%%%%%%%%%%%%%%%%%%%%%%%%%%%%%%%%%%%%% 
\begin{abstract}
The imaging and spectral performance of CdTe double-sided strip detectors (CdTe-DSDs) was evaluated for the ASTRO-H mission.
The charcterized CdTe-DSDs have a strip pitch of 0.25 mm, an imaging area of 3.2 cm$\times$3.2 cm and a thickness of 0.75 mm.
The detector was successfully operated at a temperature of $-20\mathrm{^\circ C}$ and with an applied bias voltage of 250 V.
By using two-strip events as well as one-strip events for the event reconstruction, a good energy resolution of 2.0 keV at 59.5 keV and a sub-strip spatial resolution was achieved.
The hard X-ray and gamma-ray response of CdTe-DSDs is complex due to the properties of CdTe and the small pixel effect. Therefore, one of the issues to investigate is the response of the CdTe-DSD.
In order to investigate the spatial dependence of the detector response, we performed fine beam scan experiments at SPring-8, a synchrotron radiation facility.
From these experiments, the depth structure of the electric field was determined as well as properties of carriers in the detector and successfully reproduced the experimental data with simulated spectra.
\end{abstract}

%>>>> Include a list of keywords after the abstract 


%%%%%%%%%%%%%%%%%%%%%%%%%%%%%%%%%%%%%%%%%%%%%%%%%%%%%%%%%%%%%
\section{Introduction}
\label{sec:intro}
ASTRO-H, an international X-ray satellite led by Japan scheduled for launch in 2014, aims at observing the energetic universe through energy bands ranging from soft X-rays to soft gamma rays.\cite{Takahashi2012}
Combined with the Hard X-ray Telescope (HXT), the Hard X-ray Imager (HXI), located at the focal plane of the telescope, will realize imaging spectroscopy in the hard X-ray band from 5 keV up to 80 keV.\cite{Kokubun2012}
Due to focusing, the sensitivity of the HXI for an isolated point source will be two orders of magnitude better compared with previous hard X-ray missions such as {\it INTEGRAL}\cite{INTEGRAL} or {\it Suzaku}\cite{Suzaku}.

\begin{figure}[htbp]
\centering
\includegraphics[width=11cm]{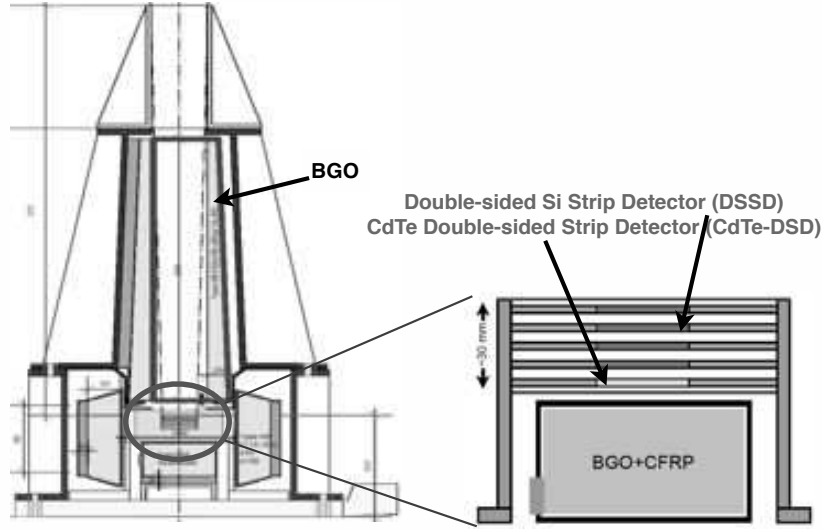}
\caption{A schematic view of the HXI (left) and the main detector (right). The main detector which is composed of four layers of double-sided Si strip detectors (DSSD) and one layer of CdTe double-sided strip detector (CdTe-DSD) is surrounded by BGO active shields.}
\label{HXI_schematic_view}
\end{figure}

\begin{figure}[htbp]
\centering
\includegraphics[width=6cm]{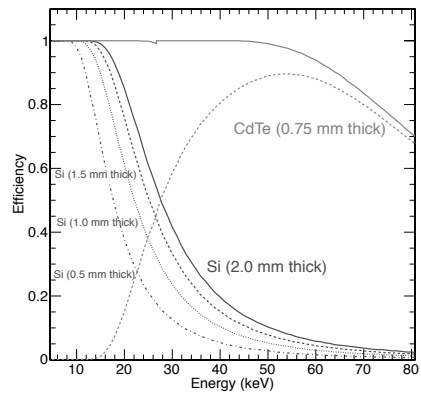}
\caption{Detection efficiency of the Si and CdTe detectors}
\label{hxi_eff}
\end{figure}

The main detector of the HXI is composed of four layers of double-sided silicon strip detectors (DSSDs) and one layer of cadmium-telluride double-sided strip detector (CdTe-DSD) as shown in the right panel of Figure \ref{HXI_schematic_view}. This hybrid structure is a result of optimization for better sensitivity throughout the observation energy range.
Figure \ref{hxi_eff} shows the detection efficiencies of the four layers of Si (thickness: 0.5 mm for each layer) and the CdTe layer (thickness: 0.75 mm). Most low-energy photons below $\sim30$ keV are detected by the DSSDs, which will be able to achieve higher sensitivity than the CdTe-DSD because of its negligible radioactivation background caused by in orbit protons.
On the other hand, the higher-energy photons up to 80 keV are detected by the CdTe-DSD with a high detection efficiency of 70--100\% due to the large atomic numbers of 48 for Cd and 52 for Te.

As the focal plane sensor of the HXT, the semiconductor imagers in the HXI have to achieve high spatial and energy resolutions.
The spatial resolution is required to be smaller than 400 $\mu$m in order to collect at least 10 data points from a image width equal to the half power diameter of the point-spread function of the telescope.
The energy resolution is required to be smaller than $\sim$2 keV (FWHM) at a photon energy of 60 keV to identify the activation background lines.
We have to evaluate these performances of the CdTe-DSD, whose designs are identical to those of the flight models, with almost same readout system as the flight models.

In addition to the measurement precision, it is important to understand the detector responses for assurance of the accuracy of the measurement, particularly for accurate interpretation of observational data.
The response of the CdTe-DSD to an X-ray photon has moderate spatial dependence because of the strip electrode configuration and ineffective charge transport properties of CdTe semiconductor.
In order to measure the spatial dependence, we performed scanning experiments of the CdTe-DSD using a photon beam at SPring-8, a synchrotron radiation facility.
In this paper, we present the imaging and spectral performances of the HXI CdTe-DSDs and development of the detector response function of CdTe-DSDs based on the scanning experiment.

%%%%%%%%%%%%%%%%%%%%%%%%%%%%%%%%%%%%%%%%%%%%%%%%%%%%%%%%%%%%%

\section{Imaging and Spectral Performance of the HXI CdTe-DSD}
\begin{figure}[htbp]
\centering
\includegraphics[width=8cm]{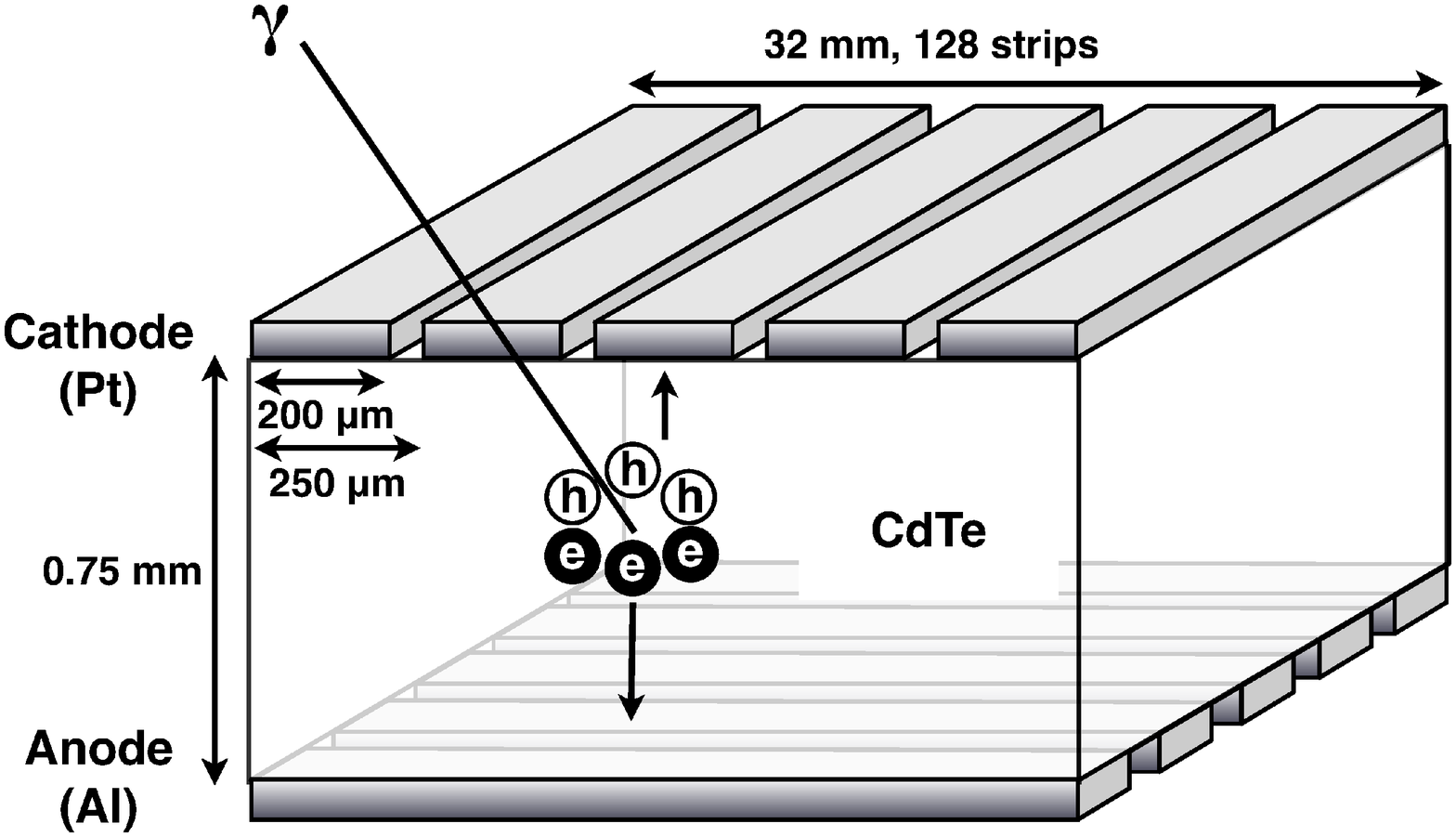}
\includegraphics[width=6cm]{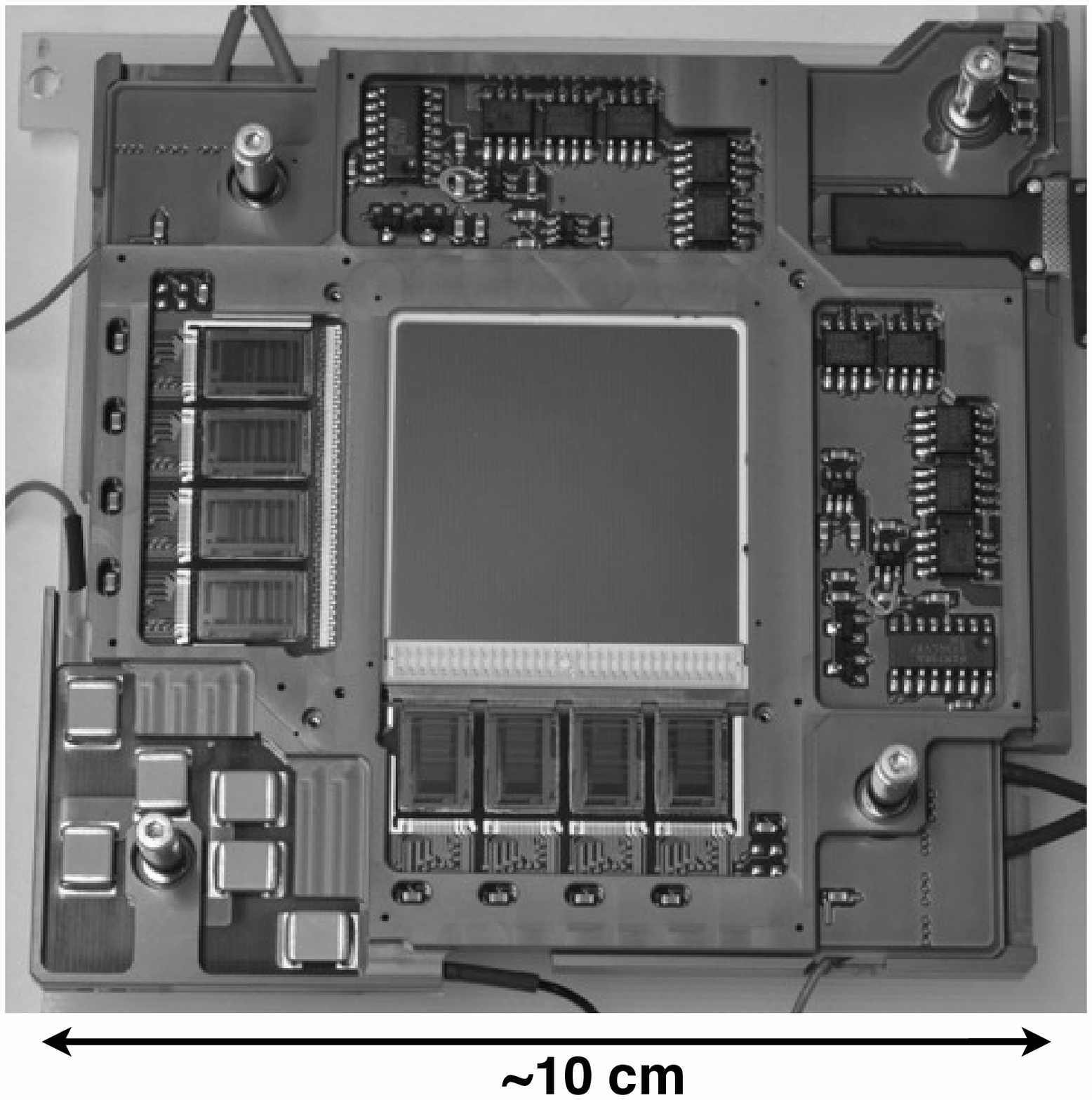}
\caption{A schematic view of the electrode configuration (left) and a photograph of the engineering model of CdTe-DSD with ASICs (right).}
\label{CdTe-DSD}
\end{figure}

The CdTe-DSDs have been developed to achieve high detection efficiency, excellent energy and spatial resolutions.\cite{Ishikawa2008,Ishikawa2010,Watanabe2009}
Table \ref{CdTe-DSD_spec} shows the specifications of our CdTe-DSD. To assure the required spatial resolution and field of view, we developed CdTe devices with a large imaging area of $32\times32$ mm$^2$ and a fine strip pitch of 250 $\mu$m, which corresponds to 0.07 arcminutes at the focal length of 12 m. Aluminum (Al) anodes and platinum (Pt) cathodes are formed on these CdTe-DSDs. Aluminum, which has a small work function, makes a Schottky barrier on the interface with a p-type CdTe crystal.\cite{Takahashi1998,Matsumoto1998,Takahashi1999,Funaki1999,Takahashi2001,Watanabe2007a,Toyama2004,Toyama2005,Watanabe2007b,Meuris2008}
Reduction of the leakage current through use of the Schottky barrier enables high energy resolution.
Figure \ref{CdTe-DSD} shows a schematic view of the electrode configuration and a photograph of an engineering model of the CdTe-DSD.
For low noise readout of the detector, we have developed a 32-channel analog LSI for the DSSD and CdTe-DSD. It is composed of trigger generation, pulse height sampling and analog-digital conversion parts. By converting to digital signals inside the ASICs, the CdTe-DSD is operated via only digital lines, and a compact and low power signal processing circuit is realized.
Kokubun et al.\cite{Kokubun2012} gives detail descriptions of the HXI CdTe-DSDs.

\begin{table}[htbp]
\caption{Specifications of CdTe-DSD\cite{Kokubun2012}}
\centering
\begin{tabular}{lll}
\hline\hline
parameter & value & note\\
\hline
electrode & Al/CdTe/Pt &\\
strip pitch & 250 $\mu$m & 0.07' @ 12 m\\
strip width & 200 $\mu$m & \\
number of strips & 128 & for each side\\
imaging area & $32\times32$ mm & FOV$\sim$9' \\
size & $34\times34$ mm & \\ 
thickness & 750 $\mu$m & \\
\hline\hline
\end{tabular}
\label{CdTe-DSD_spec}
\end{table}%

\begin{figure}[htbp]
\centering
\includegraphics[width=6cm]{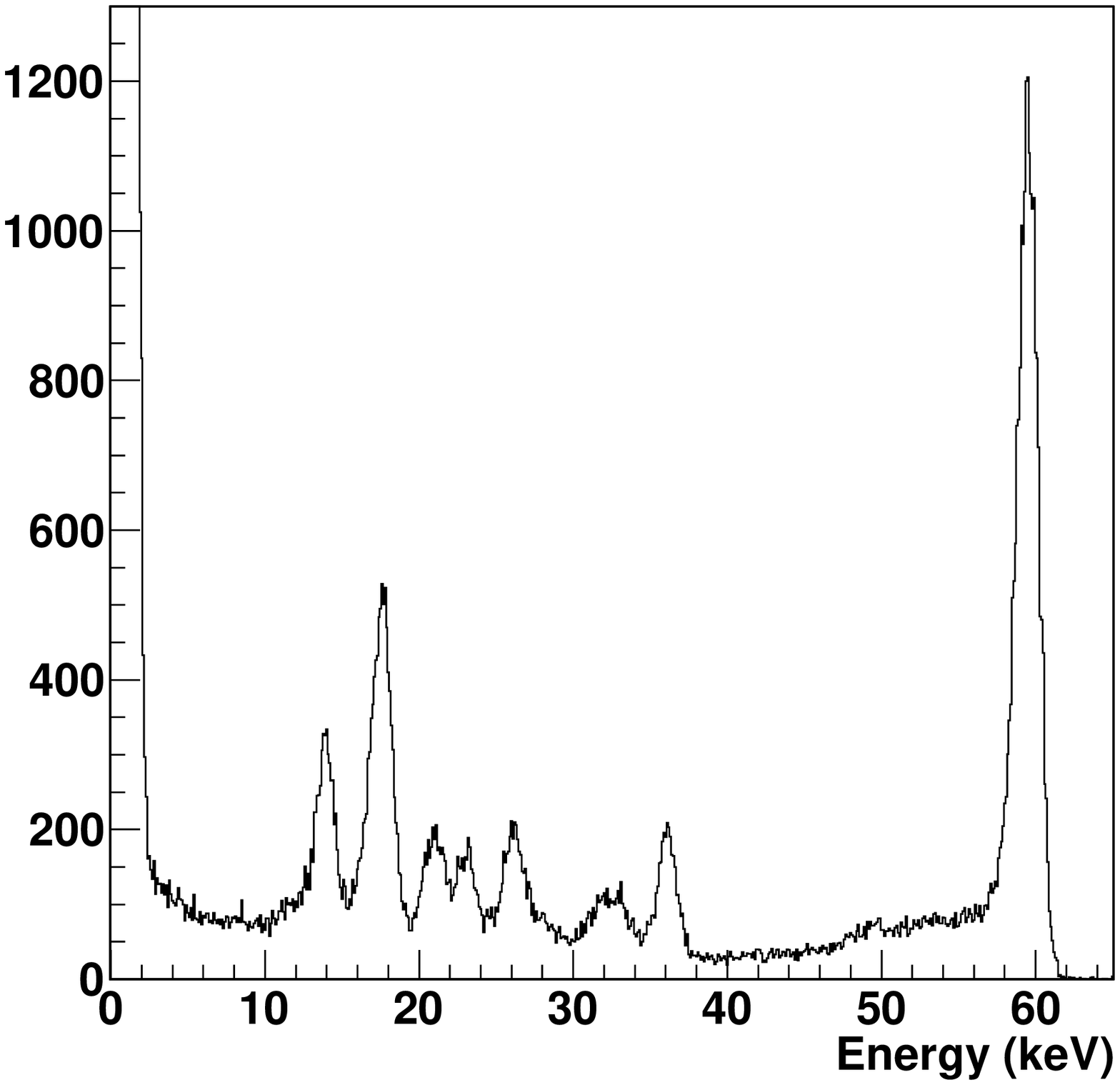}
\includegraphics[width=6cm]{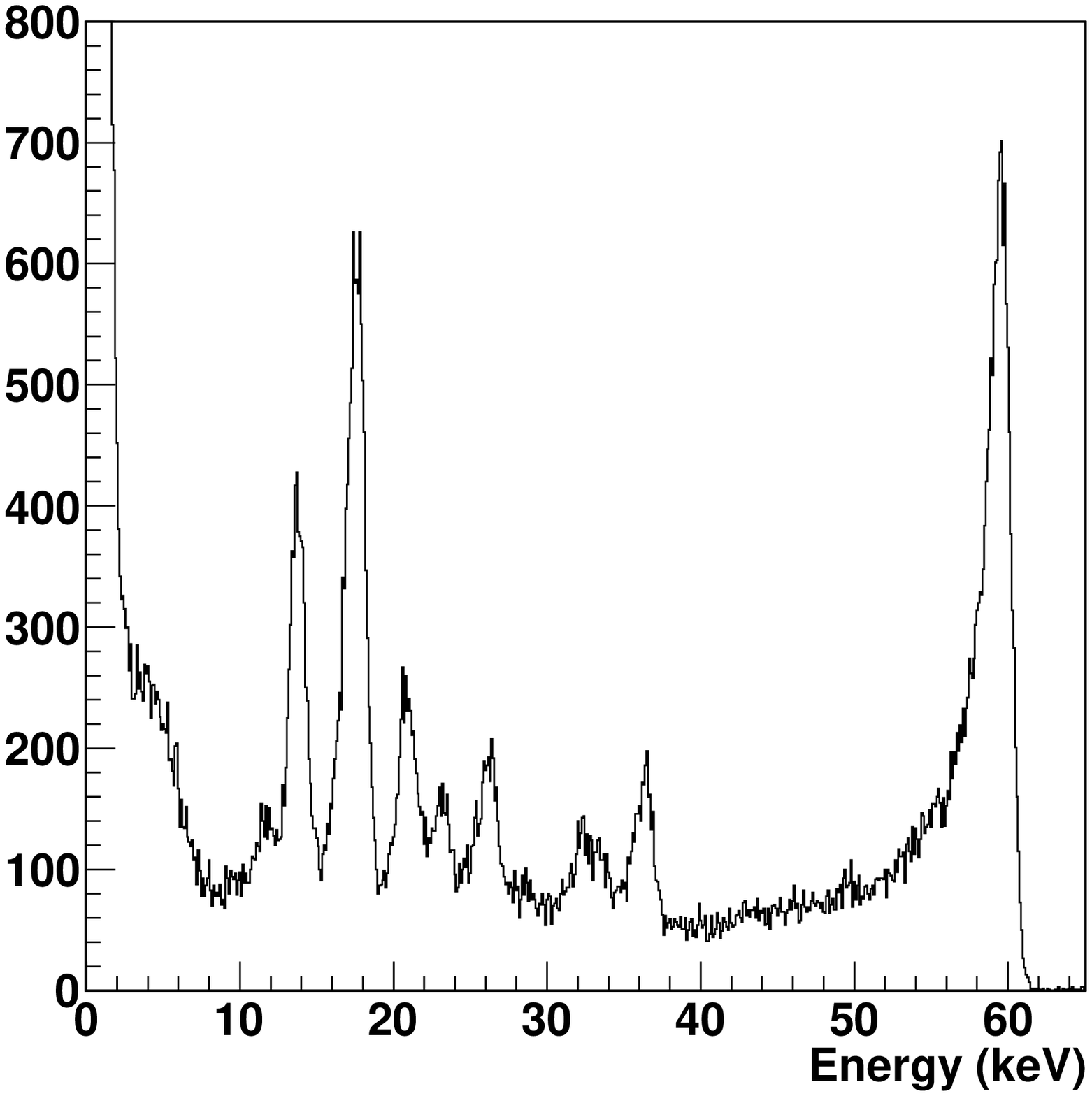}
\caption{Spectra of a $^{241}$Am source from the anode side (left) and the cathode side (right) of the CdTe-DSD. The operation temperature was $-20~\mathrm{^\circ C}$, and a bias voltage of 250 V was applied. In the best strips of each side, energy resolutions of 1.6 keV (anode) and 1.9 keV (cathode) are achieved for the 59.5 keV line.}
\label{spectra_raw}
\end{figure}

\begin{figure}[h]
\centering
\includegraphics[width=6cm]{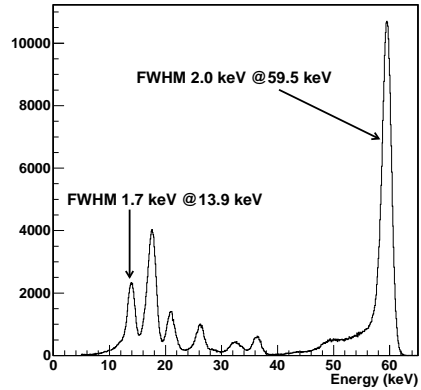}
\caption{A reconstructed spectrum of a $^{241}$Am source obtained by the CdTe-DSD for HXI. Both one-strip and two-strip events were used to make these histrograms.}
\label{spectra}
\end{figure}

The spectral performance of the CdTe-DSD was measured. 
Figure \ref{spectra_raw} shows energy spectra of a $^{241}$Am radiation source obtained from all active strips on each sides of the CdTe-DSD.
A photon interaction can be simultaneously detected in two adjacent strips; we call these events ``two-strip events". In this case, a line energy is divided into two continuum energies, or a fluorescence line and a corresponding escape line.
A large peak at 0 keV is seen due to signals from non-triggered strips.
To satisfy the requirement of the detection efficiency, it is essential to use two-strip events, which occur in about 20--30\% of all events including ``one-strip events" (those events detected at only one strip).
By removing data lower than a threshold of 5 keV and summing the signals of the two-strip events, we obtain the reconstructed spectra shown in Figure \ref{spectra}. In order to achieve the best energy resolution in the higher energy band, we adopted the energies measured at the anode side since the spectrum peak from the anode has a smaller low-energy tail structure than that of the cathode side. As a result, the target energy resolution (FWHM) of 2.0 keV at 59.5 keV has been achieved.


\begin{figure}[h]
\centering
\includegraphics[width=6cm]{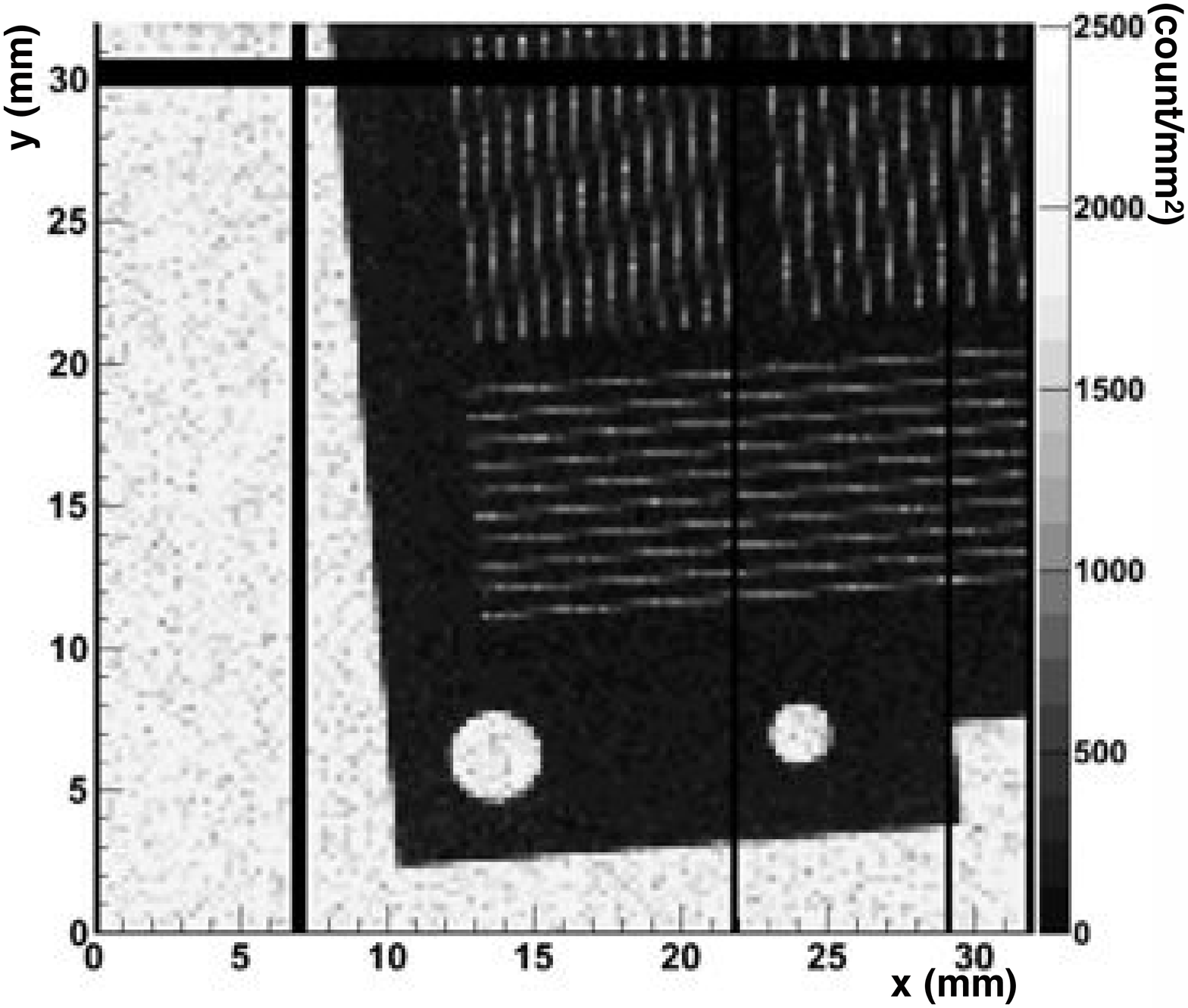}
\includegraphics[width=6cm]{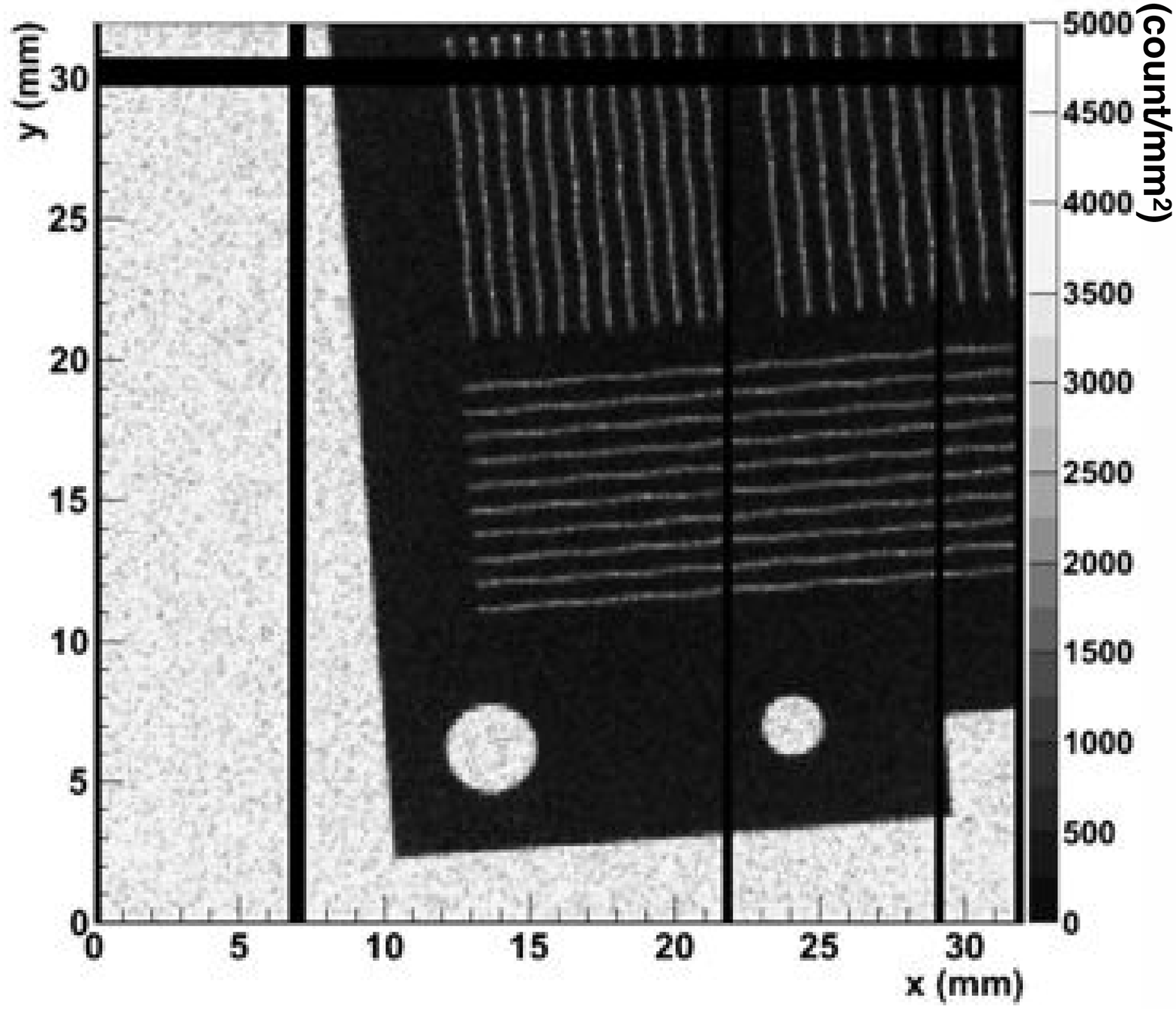}
\caption{A shadow image of a tungsten plate with 100 $\mu$m wide slits obtained with the CdTe-DSD irradiated by a radioisotope $^{241}$Am. The left image is extracted by using only one-strip events, and the right image by using both one-strip events and two-strip events.}
\label{image}
\end{figure}

Figure \ref{image} shows shadow images of a tungsten plate with 100 $\mu$m slits imaged using an X-ray energy range from 40 keV to 65 keV.
Line X-rays of 59.5 keV from $^{241}$Am are sufficiently stopped by the plate with a thickness of 0.55 mm.
In the left panel of Figure \ref{image}, only one-strip events and their interaction positions are used. The image does not clearly show the 100 $\mu$m slits because the strip pitch of the detector is 250 $\mu$m.
In the right panel, on the other hand, both the one-strip events and two-strip events were used for image reconstruction.
In the reconstruction of the two-strip events, we assumed pixels between strips. If an event was shared by two adjacent strips and neither of signals is a fluorescent energy of Cd or Te, the interaction is assumed to be within a pixel between the strips.
For other cases, the interaction is located to a pixel on the strip.
The width of the pixels is defined to be proportional to the number of events to be filled into the pixel based on the assumption of a uniformity of the sensitivity.
The experimental data of CdTe-DSD illuminated uniformly on the whole imaging area with 59.5 keV X-rays was used to define the widths.
Since the ratio of the events filled into the pixels on the strips is 74.8\% on the cathode side and 82.4\% on the anode side in the data, the width of the pixel was defined as 187 $\mu$m on the cathode strip and 206 $\mu$m on the anode strip. These values do not contradict the results of the scanning experiments.
As a result of these event reconstructions, in contrast to the left panel, the 100 $\mu$m slits are apparent in the right panel image.
Using two-strip events enables sub-strip spatial resolution.

%%%%%%%%%%%%%%%%%%%%%%%%%%%%%%%%%%%%%%%%%%%%%%%%%%%%%%%%%%%%%
\section{Modeling of Detector Response}
For the accurate interpretation of the X-ray data, accurate knowledge on detector response is of great importance.
In order to model the detector response, we have developed a full Monte Carlo simulator.\cite{Odaka2010}
It treats particle tracking, signal generation in the detector devices, and signal processing including data acquisition system. For particle tracking, we use Monte Carlo simulation based on the Geant4 toolkit\cite{Allison2006} to obtain energy deposits and interaction positions in CdTe material. By multiplying the energy deposited by charge collection efficiency (CCE) at the interaction position, the signal induced on the electrode can be calculated. 

Due to ineffective charge transport properties of CdTe detectors, holes and electrons (carriers) generated by incident photons are not fully collected.
Charge induced on electrodes can be calculated by the Shockley-Ramo theorem.\cite{He2001,Zumbiehl2001,Eskin1999}
In this theorem, a mathematical concept of weighting potential $\phi_w$ plays an important role in the calculation. 
When charge $q(\mbox{\boldmath $x$})$ moves from an interaction position $\mbox{\boldmath $x$}_i$ to a position $\mbox{\boldmath $x$}_f$, induced charge $Q$ at an electrode can be calculated by
\begin{eqnarray}
Q=\int_{\mbox{\boldmath $x$}_i}^{\mbox{\boldmath $x$}_f}q(\mbox{\boldmath $x$})\mbox{\boldmath $E$}_w(\mbox{\boldmath $x$})\cdot d\mbox{\boldmath $x$}\label{wp_to_Q}
\end{eqnarray}
where $\mbox{\boldmath $E$}_w(\mbox{\boldmath $x$})\equiv -\mbox{\boldmath $\nabla$}\phi_w(\mbox{\boldmath $x$})$ is a weighting field.
Due to the finite lifetime $\tau$ of the carrier, charge $q$ decreases exponentially as $q(t)=q_0\exp(-t/\tau)$.
The time $t$ can be related to the position $\mbox{\boldmath $x$}$ by the equation $d\mbox{\boldmath$x$}/dt=\mbox{\boldmath$v$}=\mu\mbox{\boldmath$E$}(\mbox{\boldmath$x$})$, where $\mbox{\boldmath$E$}(\mbox{\boldmath$x$})$ is an electric field (not weighting field), $\mbox{\boldmath$v$}$ is the drift velocity and $\mu$ is the mobility of carriers.
Therefore, the CCE, defined as $Q/q_0$, is a function of the interaction position $\mbox{\boldmath $x$}_i$.

\begin{figure}[htbp]
\centering
\includegraphics[width=8cm]{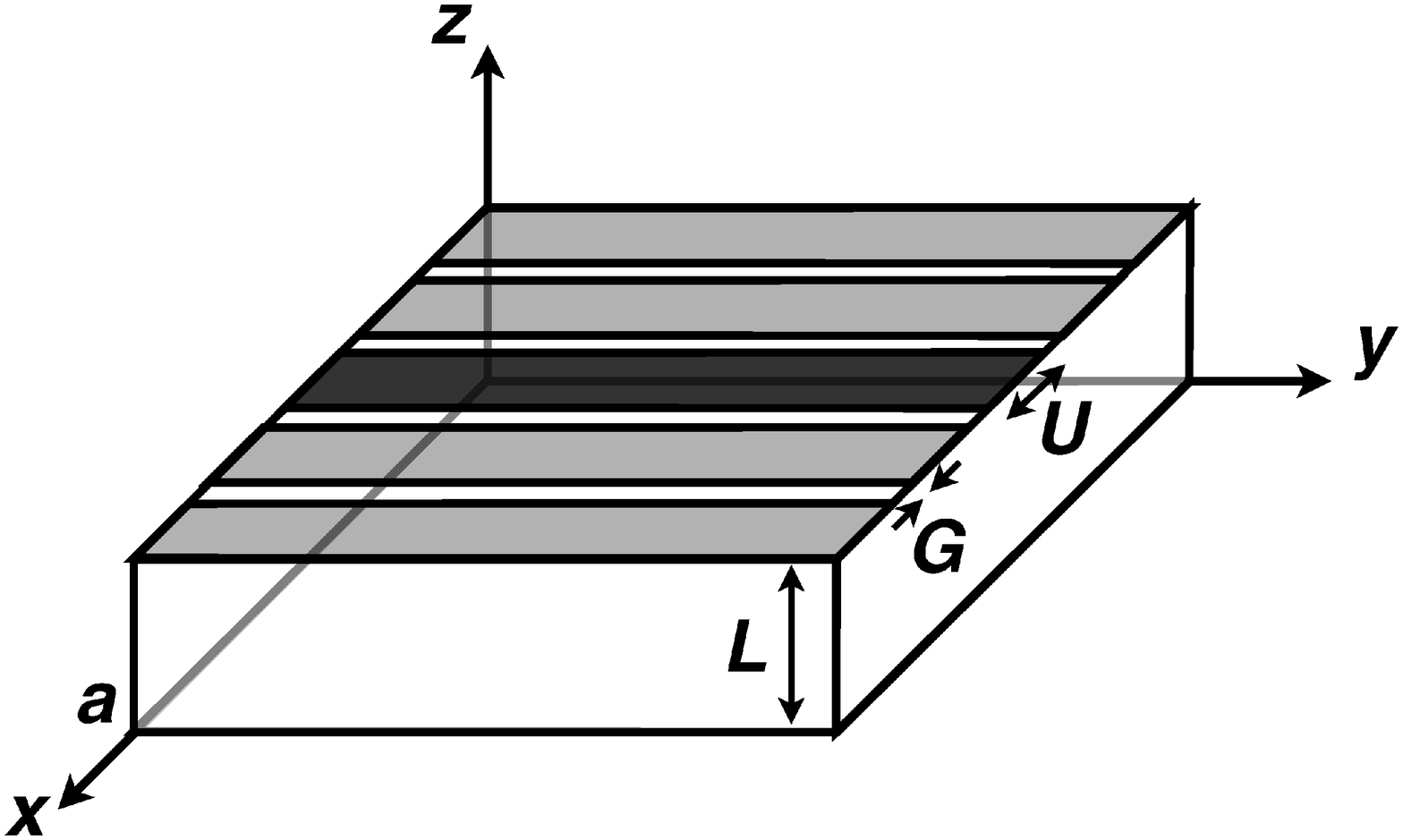}
\includegraphics[width=6cm]{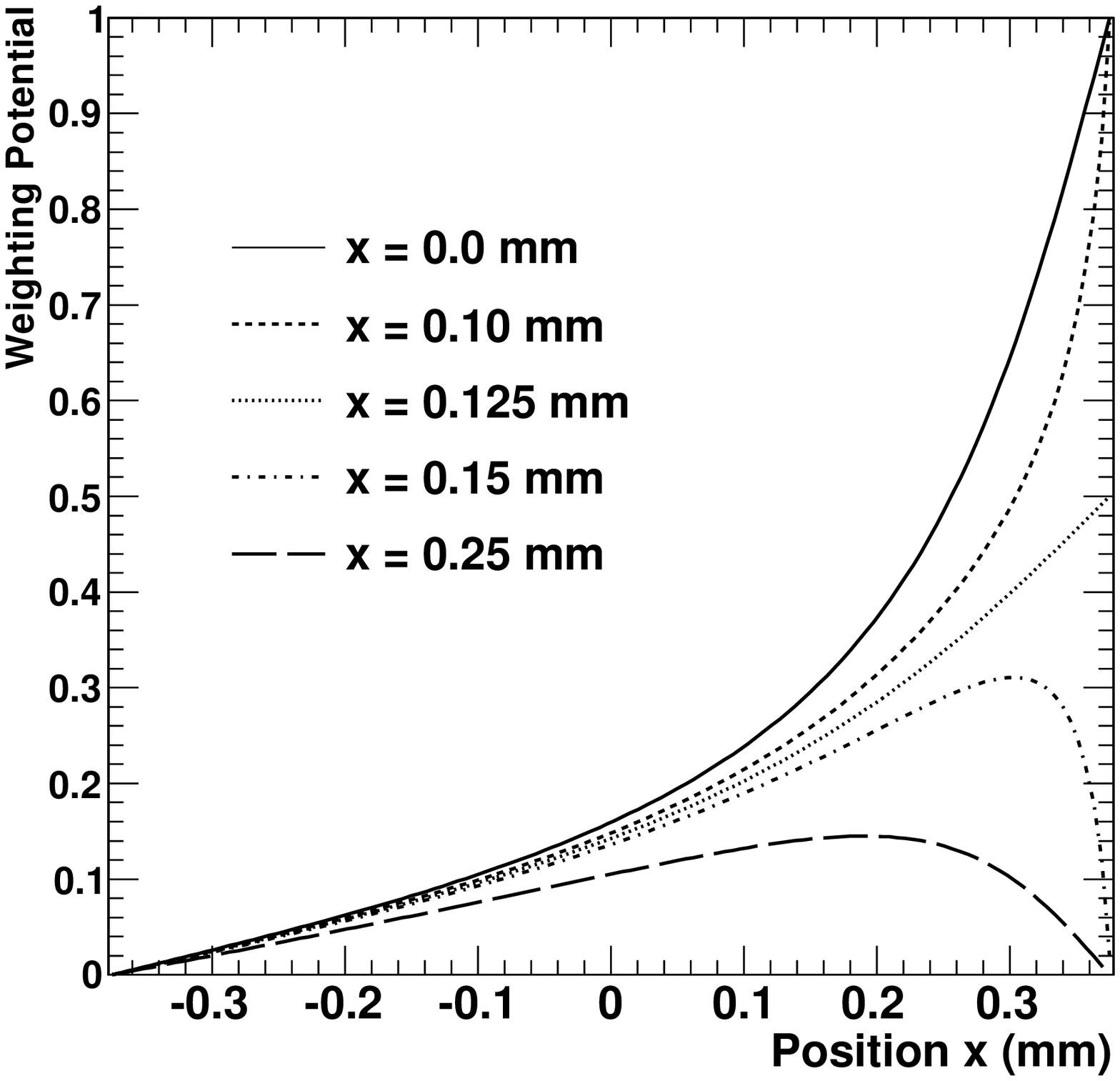}
\caption{The left panel shows definitions of geometrical parameters for the calculation of the weighting potential in a strip detector. The right panel shows the weighting potentials in the HXI CdTe-DSD as a function of depth. The solid line (the center of the readout electrode), dashed line (the boundary between the readout electrode and the gap), dotted line (the center of the gap), dot-dashed line (the boundary between the gap and the next electrode), long-dashed line (the center of the next electrode) are drawn separately for indicating spatial dependence in the horizontal direction.}
\label{wp_config}
\end{figure}

The weighting potential is calculated by solving Poisson's equation $\Delta\phi_w(\mbox{\boldmath $x$})=0$ with the boundary condition of $\phi_w=1$ at the readout electrode and $\phi_w=0$ at all the other electrodes.
The weighting potential in the strip detector as shown in the left panel of Figure \ref{wp_config} is calculated by
\begin{eqnarray}
\phi_w(x,z)&=&\sum_{m=1}^\infty A_m\sin\left( \alpha_mx\right)\sinh\left( \alpha_mz\right)\label{wp}\\
A_m&=&\frac{2}{\alpha_m^2Ga\sinh(\alpha_m L)}f_m\\
f_m&=&\sin\left( \frac{\alpha_m}{2}(a-U)\right)-\sin\left( \frac{\alpha_m}{2}(a-U-2G)\right)\nonumber\\
&&+\sin\left( \frac{\alpha_m}{2}(a+U)\right)-\sin\left( \frac{\alpha_m}{2}(a+U+2G)\right) \label{wp_coefficient}
\end{eqnarray}
where $\alpha_m=m\pi/a$, $a$ is the detector size, $U$ is the width of the electrodes, $G$ is the width of the inter-electrode gaps, and $L$ is the thickness of the detector (see the left panel of Figure \ref{wp_config}). For simplicity, we assumed that the boundary condition at the gap between two electrodes is linearly interpolated from the values at these electrodes. The right panel of Figure \ref{wp_config} shows the weighting potential in the HXI CdTe-DSD, where the detector size $a=34$ mm, the thickness $L=0.75$ mm, the width of the electrode $U=0.20$ mm and the width of the gap $G=0.05$ mm.
Since the weighting potential is distorted due to the strip electrode configuration, the motion of holes mainly contributes to a signal induced on the cathode, and the motion of electrons to the anode, namely, the small pixel effect.
Since the value of $\mu\tau$ of holes is one order of magnitude smaller than that of electrons in CdTe, depth dependence of detector response is enhanced by this effect.

Moreover, crosstalk caused by the weighting potential increases the complexity of detector response. As seen in the right panel of Figure \ref{wp_config}, the weighting potential in regions of the strip next to the readout strip has a considerably higher value.
This means that if the carriers are trapped before reaching the electrodes, a significant signal should be induced in a strip next to the strip closest to the interaction.
In addition to the crosstalk, thermal diffusion and Coulomb repulsion of carriers affect the position dependence of the detector response. In the simulation, the effect of the diffusion and Coulomb repulsion are implemented by spreading carriers for every small time interval $\Delta t$. The structure of the electric field, which has not been understood or measured sufficiently, also affects the drift velocity and the path of carriers in the CdTe detector.

%%%%%%%%%%%%%%%%%%%%%%%%%%%%%%%%%%%%%%%%%%%%%%%%%%%%%%%%%%%%%
\section{Beam Scanning Measurement} \label{sec:sections}
\begin{figure}[htbp]
\centering
\includegraphics[width=11cm]{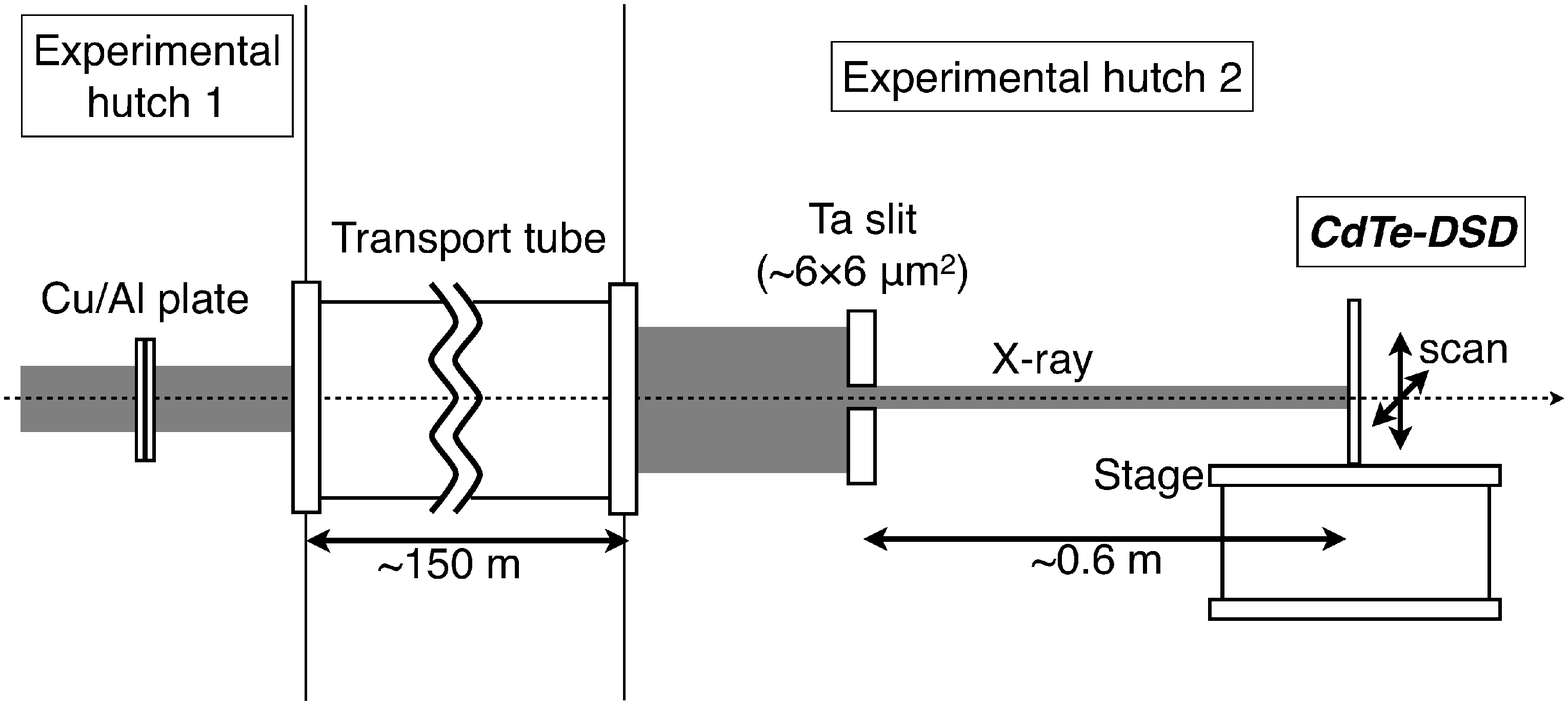}
\includegraphics[width=5cm]{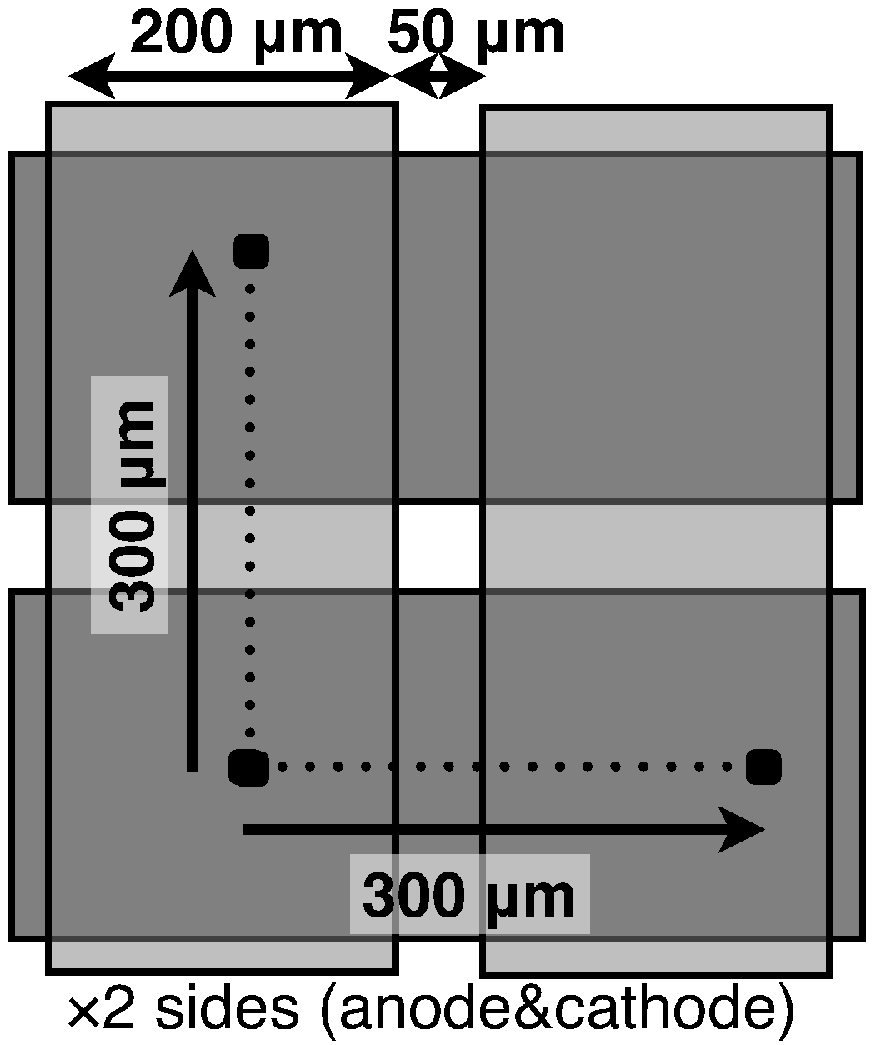}
\caption{A schematic view of the experimental setup at the BL20B2 in SPring-8 (left panel) and scanning procedures (right panel).}
\label{SP8exp}
\end{figure}

As discussed in the previous section, it is important to understand spatial dependence of the detector response.
For this purpose, a scanning experiment of the HXI CdTe-DSD was performed at the beamline BL20B2 of the synchrotron radiation facility SPring-8 in Hyogo, Japan.
BL20B2 is a 215 m long beamline developed for medical imaging research.\cite{Goto2001}
A monochromatic X-ray beam was used, extracted from synchrotron X-rays by a double-crystal monochromator at a distance of 36.8 m from the bending magnet source. In our experiment, the Si (111) diffraction was chosen to obtain 30 keV photons. Its energy uncertainty was $\Delta E/E\sim10^{-4}$, which is sufficiently smaller than the energy resolution of the detectors ($\Delta E/E\sim10^{-2}$--$10^{-1}$).
Figure~\ref{SP8exp} shows the experimental setup.
The monochromatic X-ray bream was attenuated before the CdTe detector by a 0.5 mm thick Al plate and a 0.5 mm thick Cu plate at Hutch 1 (44 m from the source), and then collimated by a tantalum slit at Hutch 2 (203 m from the source).
At 60 m away from the slit, the CdTe-DSD was located on a movable stage. Both sides of the CdTe-DSD were scanned along the two axes with 10 $\mu$m steps over 300 $\mu$m.
During the experiment, the detector was being operated at a temperature of approximately $-14\mathrm{^\circ C}$.
\begin{table}[htbp]
\caption{Experimental conditions}
\centering
\begin{tabular}{cc}
\hline\hline
X-ray energy & 30 keV (and 90 keV etc.)\\
Beam size & 6$\times$6 $\mathrm{\mu m^2}$ (slit aperture)\\
Scan pattern & 2 axes$\times$2 side\\
Bias & 250 V\\
Temperature & $-14\mathrm{^\circ C}$\\
Scan length & 300 $\mu$m (31 spot)\\
\hline\hline
\end{tabular}
\label{exp_cond}
\end{table}%

\begin{figure}[htbp]
\centering
\includegraphics[width=6cm]{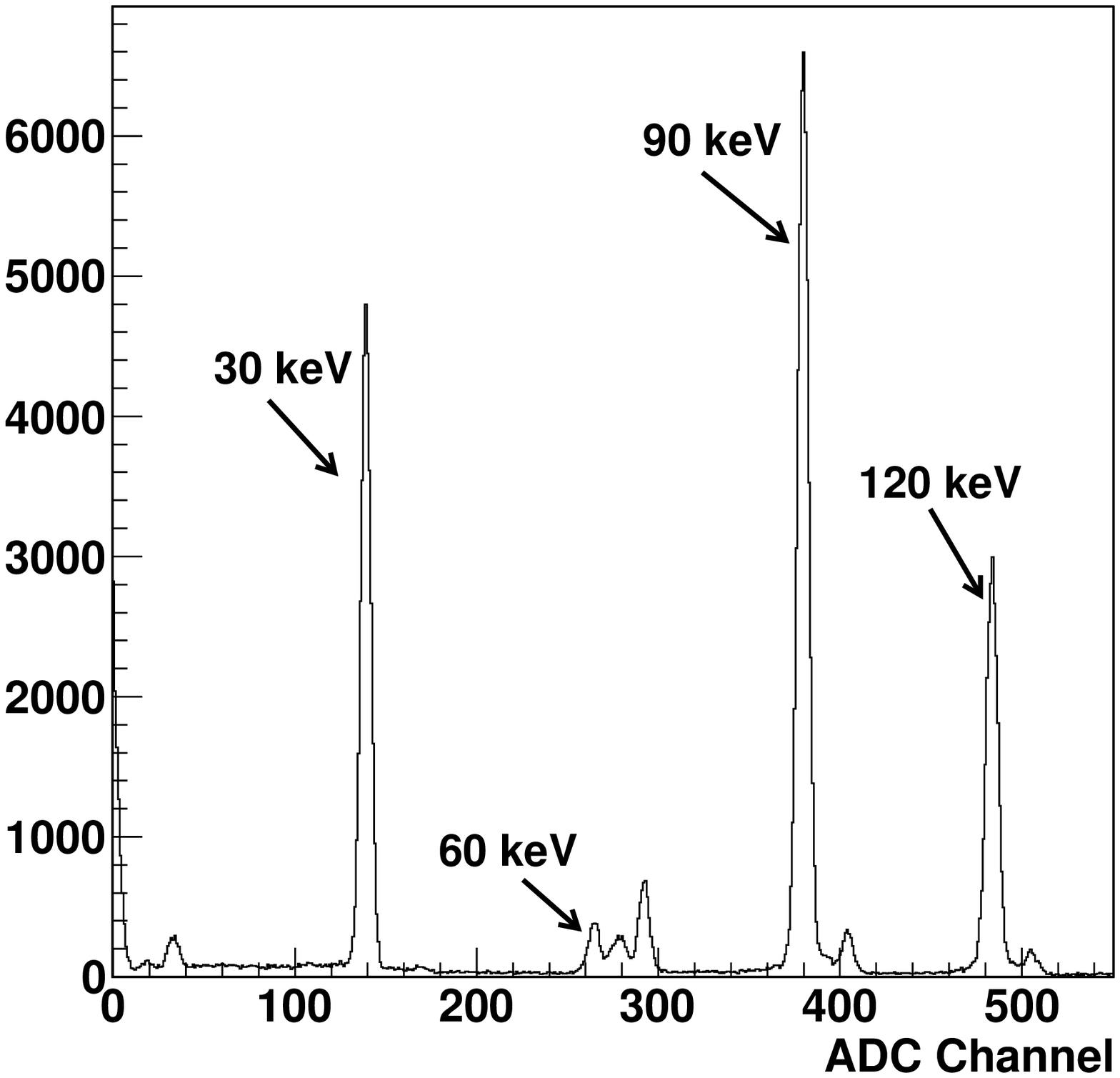}
\includegraphics[width=7cm]{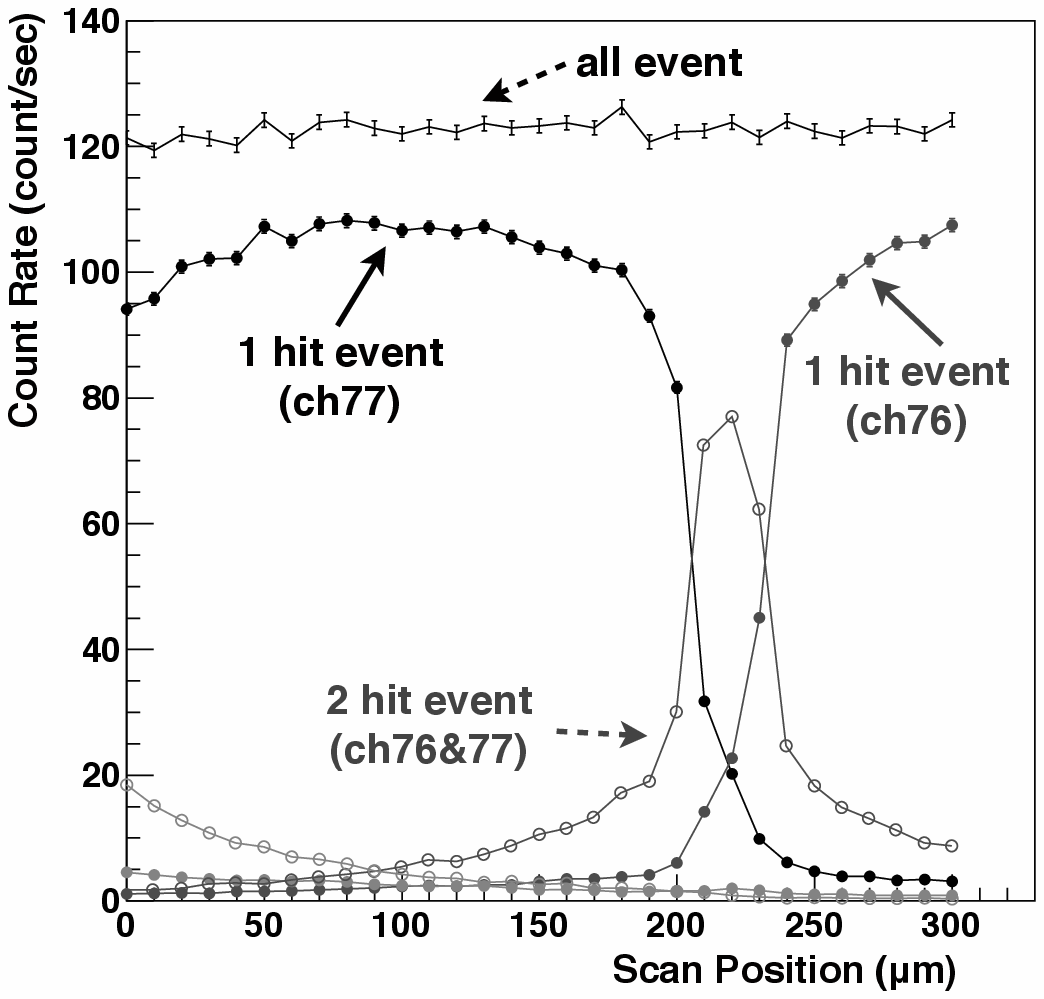}
\caption{The left panel shows an example of anode side spectrum obtained by the scanning experiment at SPring-8. The lines at a little higher energy than 60 keV etc. are the fluorescence lines. The right panel shows count rates in an energy range of 20 keV to 40 keV for different event types. The standard deviation of the count rate of all the events is about 1\%, which is consistent with the statistical errors.}
\label{scan_result}
\end{figure}

A spectrum obtained by the scanning experiment is shown in the left panel of Figure \ref{scan_result}.
Since the monochromatic X-rays are separated by Bragg reflections, higher harmonics lines are also seen in the spectrum. The energy resolution (FWHM) on the anode side, which is mainly determined by the readout noise, was 1.4 keV at 30 keV, and 1.9 keV at 90 keV. These correspond to the FWHM in ADC values of 6.7 at 30 keV, and 7.2 at 90 keV.
The values of the energy resolution were consistent with those obtained in the performance tests in our laboratory, assuring quality of the data for investigating the detector response.
The right panel of Figure \ref{scan_result} shows count rates in an energy range from 20 keV to 40 keV. The count rate of the two-strip events increases at the electrode gap ($\sim220~\mu$m). Due to the weighting potential crosstalk (as seen in Section 3) and the fluorescence lines escaping to the next strip, the two-strip events exist at the center of electrode ($\sim100~\mu$m), and smoothly increase toward the gap.

\begin{figure}[htbp]
\centering
\includegraphics[width=6cm]{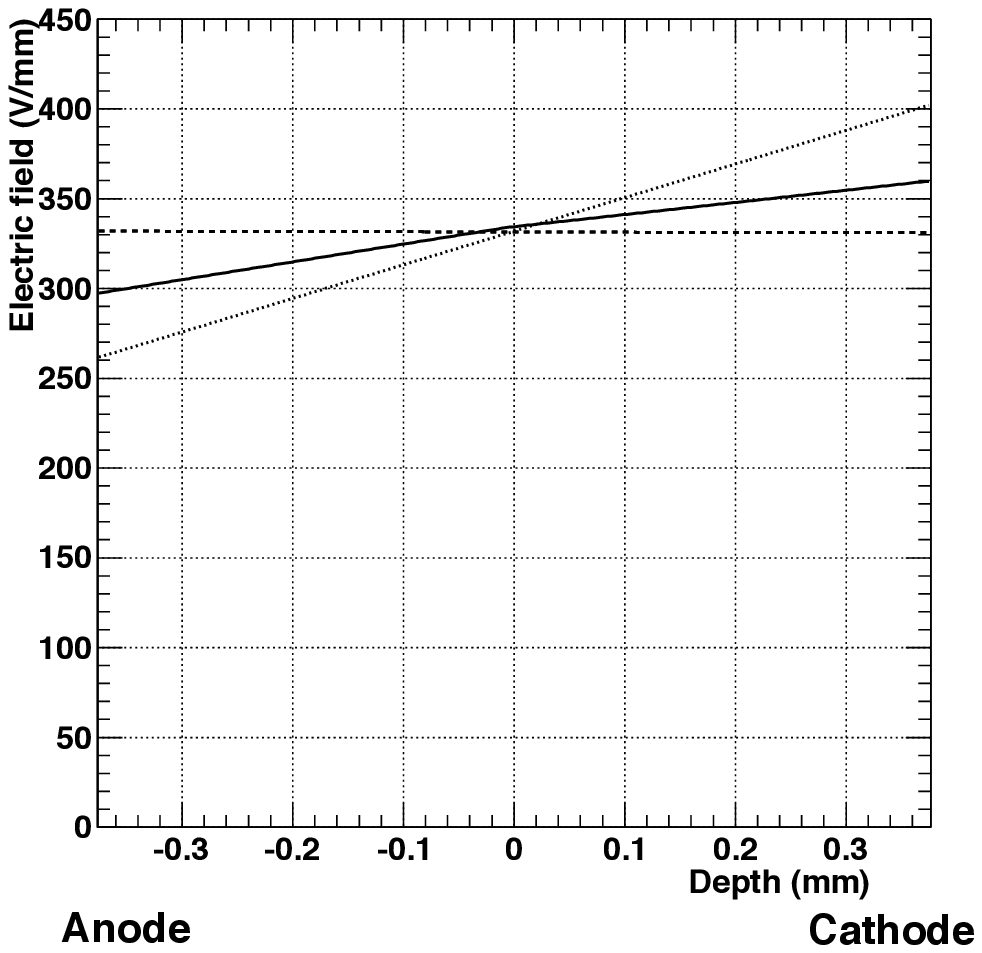}
\includegraphics[width=6cm]{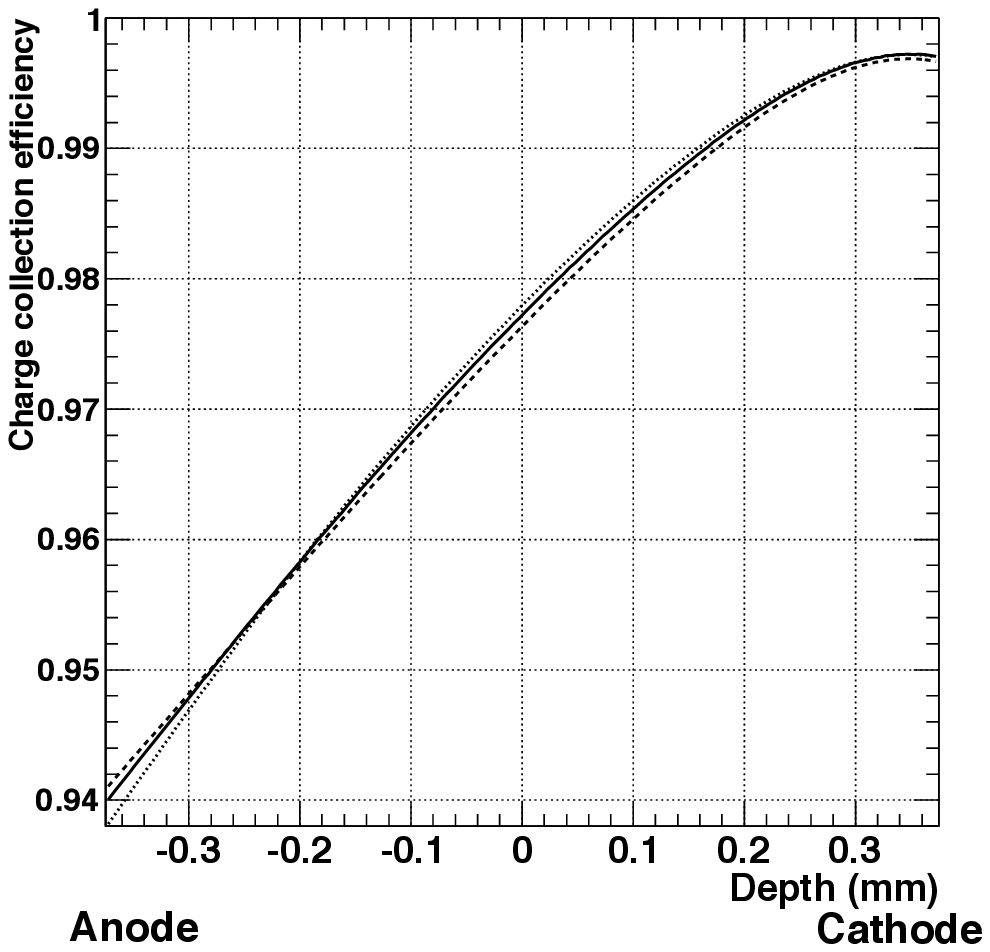}
\includegraphics[width=6cm]{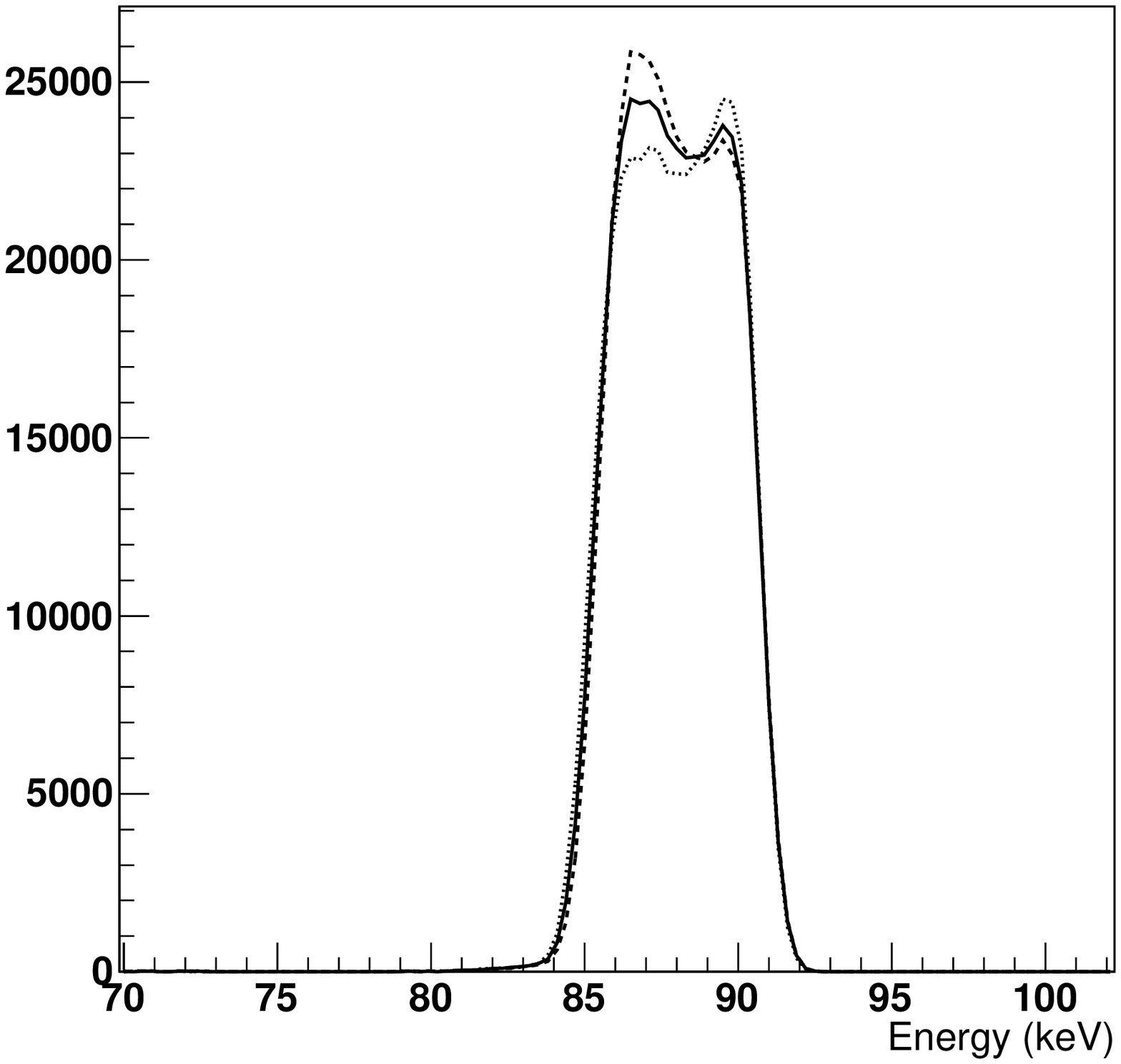}
\caption{The assumed electric fields (upper left), the charge collection efficiencies (upper right), and simulated cathode side spectra of anode irradiation (lower) for different electric fields.}
\label{efz}
\end{figure}

The depth dependence of the electric field affects the cathode side spectrum irradiated on the center of the electrode from the anode side as shown in Figure \ref{efz}. Each line corresponds to different space charge densities in CdTe of $\sim0~\mathrm{cm^{-3}}$ (dashed lines), $\sim5.5\times10^{10}~\mathrm{cm^{-3}}$ (solid lines) and $\sim1\times10^{11}~\mathrm{cm^{-3}}$ (dotted lines).
Each model spectrum shows two peaks. The lower peak is generated by the events interacting near the anode and the higher peak by the events interacting near the cathode.
If the electric field were stronger near the cathode than near the anode (dotted lines), the charge collection efficiency around the middle of CdTe would tend to be higher.
Therefore the higher peak has more events than the lower peak when the positive space charge exists in CdTe.

\begin{figure}[htbp]
\centering
\includegraphics[width=14cm]{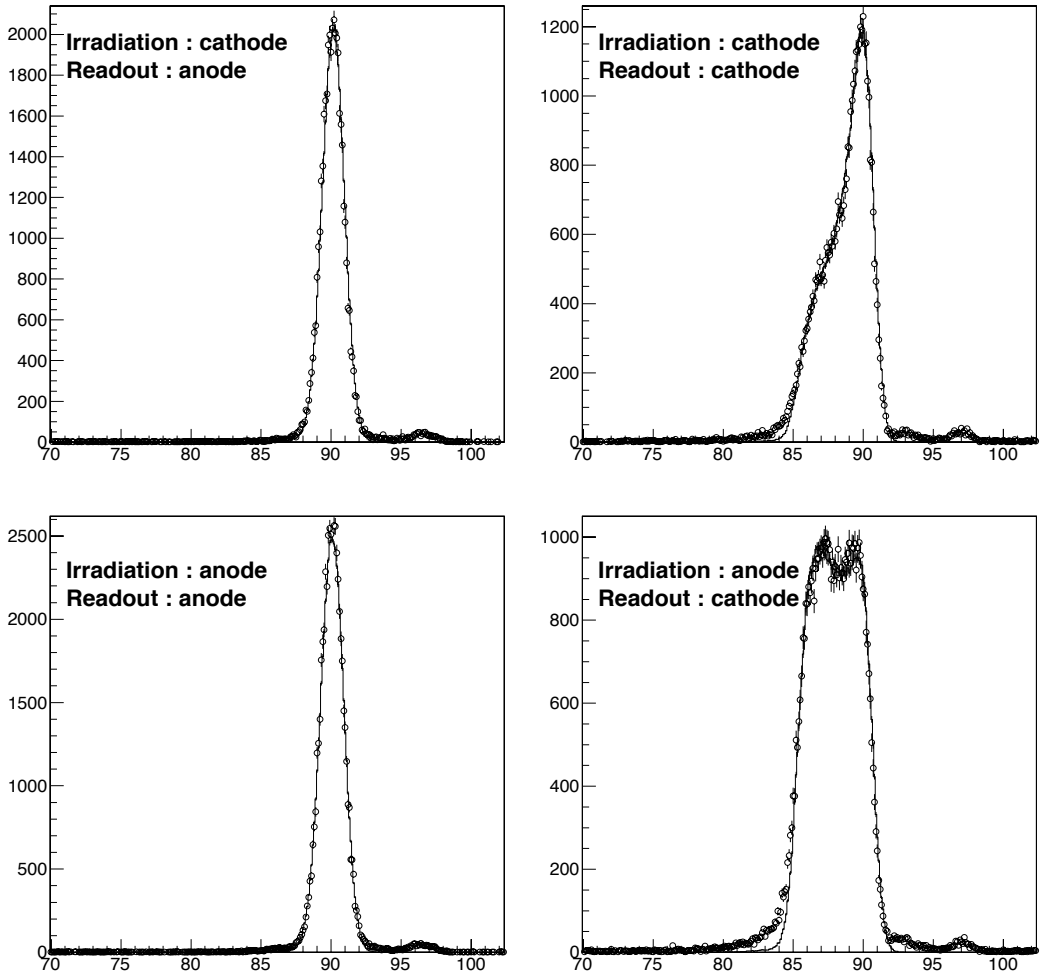}
\caption{Simulated spectra (solid line) and experimental spectra (circle) for different irradiation side and readout side. The lines above 92 keV in the experimental spectra is the fluorescence lines.}
\label{compare_specs}
\end{figure}

Spectra of the CdTe-DSD were simulated and compared with the experimental spectra by using the model including the depth distribution of the electric field shown in the above discussion. 
In order to determine the parameters in the CdTe detector, the value of $\chi^2$ between the experimental spectra and the simulated spectra was calculated and minimized. In the minimization, we used the SIMPLEX method in the MINUIT library.\cite{MINUIT}
The best fit parameters are shown in Table \ref{parameters}.
The electric field shown by a solid line in Figure \ref{efz} is the best fit model. Therefore the electric field in the CdTe-DSD must increase from the anode. The same property of the electric field in p-type CdTe detectors is reported by Zumbiehl et al. by measurement based on Pockels effect.\cite{Zumbiehl1999}
Figure \ref{compare_specs} shows the simulated and experimental spectra of the one-strip events irradiated on the center. The experimental spectra was successfully reproduced with the detector model, though tail structures below 85 keV in the spectra of cathode side have small disagreements, probably due to the weak electric field near the gap between the electrodes.
\begin{table}[htbp]
\caption{Model prameters}
\centering
\begin{tabular}{ll}
\hline\hline
$(\mu\tau)_{\mathrm{hole}}$ & $2.81\times10^{-4}~\mathrm{cm^2/V}$\\
$(\mu\tau)_{\mathrm{electron}}$ & $1.82\times10^{-3}~\mathrm{cm^2/V}$\\
space charge density & $5.5 \times 10^{10}~\mathrm{cm^{-3}}$\\
noise at 90 keV (anode) & 1.78 keV\\
noise at 90 keV (cathode) & 1.43 keV\\
\hline\hline
\end{tabular}
\label{parameters}
\end{table}%

%%%%%%%%%%%%%%%%%%%%%%%%%%%%%%%%%%%%%%%%%%%%%%%%%%%%%%%%%%%%%
\section{Summary}
The spectral and imaging performance of the CdTe double-sided strip detectors (CdTe-DSDs) was evaluated for the HXI.
The detector was successfully operated at a temperature of $-20\mathrm{^\circ C}$ and a bias voltage of 250 V.
After event reconstructions, a energy resolution of 2.0 keV (FWHM) at 59.5 keV and a spatial resolution $\le$ 250 $\mu$m pitch was achieved. This performance is sufficient for the HXI.
In order to investigate the spatial dependence in detector response, scanning experiments at SPring-8, a synchrotron radiation facility, were performed.
By comparing the simulated spectra and the experimental spectra obtained by the scanning experiment, a model including the depth dependences of electric field was constructed, allowing determination of the key characteristics of the CdTe-DSD such as charge transport properties and the electric field structure.
Under the assumption of a positive space charge with a uniform density of $\sim5.5\times10^{10}~\mathrm{cm^{-3}}$, the experimental spectra were successfully reproduced using the detector model.

%%%%%%%%%%%%%%%%%%%%%%%%%%%%%%%%%%%%%%%%%%%%%%%%%%%%
%\appendix    %>>>> this command starts appendixes
%%%%%%%%%%%%%%%%%%%%%%%%%%%%%%%%%%%%%%%%%%%%%%%%%%%%
%\section{SPring-8} \label{sec:misc}

%%%%%%%%%%%%%%%%%%%%%%%%%%%%%%%%%%%%%%%%%%%%%%%%%%%%%%%%%%%%%
%\acknowledgments     %>>>> equivalent to \section*{ACKNOWLEDGMENTS}       
 
%%%%%%%%%%%%%%%%%%%%%%%%%%%%%%%%%%%%%%%%%%%%%%%%%%%%%%%%%%%%%
%%%%% References %%%%%

\bibliography{report}   %>>>> bibliography data in report.bib
\bibliographystyle{spiebib}   %>>>> makes bibtex use spiebib.bst

\end{document}